%% file: pmain.tex
\newif\ifdraft
\newif\ifcomment
\newif\iflatexdiff
\newif\ifpreprint
\def\dvers{v1.0}
\def\dtitle{Jet-like correlations with neutral pion triggers\\ in pp and central Pb--Pb collisions at 2.76 TeV} 
\def\stitle{Jet-like correlations with neutral pion triggers at 2.76 TeV} 
\documentclass[ALICE,manyauthors,12pt]{cernphprep}
\usepackage[comma,square,numbers,sort&compress]{natbib}
\usepackage{hyperref}
\usepackage{graphicx}  
\usepackage{dcolumn}   
\usepackage{bm}        
\usepackage{amssymb}   
\usepackage{amsfonts}
\usepackage{graphics}
\usepackage{grffile}   
\usepackage{epsfig}
\usepackage{units}
\usepackage[usenames]{color}
\usepackage[normalem]{ulem} 
\usepackage{color}
\usepackage[utf8]{inputenc}
\usepackage[T1]{fontenc}
\usepackage{subfigure}
\usepackage{multirow}
\usepackage[utf8]{inputenc}
\iflatexdiff
\RequirePackage{color}\definecolor{RED}{rgb}{1,0,0}\definecolor{BLUE}{rgb}{0,0,1}

\fi

\input{commands.tex}

\graphicspath{{./img/}}
\usepackage{lineno}
\ifdraft
\linenumbers
\modulolinenumbers[1]
\usepackage{fancyhdr}
\fi
\begin{document}
\begin{titlepage}
\PHyear{2016}
\PHnumber{195}                   
\PHdate{02 Aug}                  
\title{\dtitle}
\ShortTitle{\stitle}
\Collaboration{ALICE Collaboration%
         \thanks{See Appendix~\ref{app:collab} for the list of collaboration members}}
\ShortAuthor{ALICE Collaboration} 
\ifdraft
\begin{center}
 \today\\ \color{red}DRAFT \dvers\ \hspace{0.3cm} \$Revision: 2904 $\color{white}:$\$\color{black}\vspace{0.3cm}
\end{center}
\fi
\begin{abstract}
\input{abstract.tex}
\end{abstract}
\end{titlepage}
\newpage
\setcounter{page}{2}

\input{pcontent.tex}

\newenvironment{acknowledgement}{\relax}{\relax}
\begin{acknowledgement}
\section*{Acknowledgements}
We thank Hanzhong Zhang and Guo-Liang Ma for providing the AMPT and pQCD predictions, respectively.

\input{acknowledgements.tex}
\end{acknowledgement}
\bibliographystyle{utphys}
\bibliography{biblio}{}
\newpage
\appendix
\section{The ALICE Collaboration}
\label{app:collab}
\input{Alice_Authorlist_2016-Aug-01.tex}

\end{document}

%% file: commands.tex
\newcommand{\piz}          {\ensuremath{\pi^{0}}}
\newcommand{\pt}           {\ensuremath{p_{\mathrm{T}}}}
\newcommand{\kt}           {\ensuremath{k_{\mathrm{T}}}}
\newcommand{\ptt}          {\ensuremath{p_{\mathrm{T}}^{\rm trig}}}
\newcommand{\pta}          {\ensuremath{p_{\mathrm{T}}^{\rm assoc}}}
\newcommand{\qhat}         {\ensuremath{\hat{q}}}
\newcommand{\gmom}         {\mathrm{GeV}/c}
\newcommand{\dphi}         {\ensuremath{\Delta\varphi}}
\newcommand{\deta}         {\ensuremath{\Delta\eta}}

\newcommand{\snn}          {\ensuremath{\sqrt{s_{\mathrm{NN}}}}}
\newcommand{\Iaa}          {\ensuremath{I_{\mathrm{AA}}}}

\newcommand{\PbPb}         {\mbox{Pb--Pb}}

\newcommand{\AuAu}         {\mbox{Au--Au}}

\newcommand{\av}[1]        {\ensuremath{\left\langle#1\right\rangle}}
\newcommand{\NLM}          {\ensuremath{N_{\rm LM}}}
\newcommand{\lzt}          {\ensuremath{\sigma^2_{\rm long}}}
\newcommand{\lztmin}       {\ensuremath{\lambda_{\rm min}}}
\newcommand{\lztmax}       {\ensuremath{\lambda_{\rm max}}}
\newcommand{\dd}           {\ensuremath{{\rm d}}}
\newcommand{\Ntrig}        {\ensuremath{N_{\rm trig}}}
\newcommand{\Nassoc}       {\ensuremath{N_{\rm assoc}}}
\newcommand{\Fig}[1]       {Fig.~\ref{#1}}
\newcommand{\Figure}[1]    {Figure~\ref{#1}}

\newcommand{\Section}[1]   {Section~\ref{#1}}

\newcommand{\Eq}[1]        {Eq.~\ref{#1}}

\newcommand{\Ref}[1]       {Ref.~\cite{#1}}

\newcommand{\Tab}[1]       {Tab.~\ref{#1}}

\newcommand{\co}[1]        {\relax}

%% file: abstract.tex
We present measurements of two-particle correlations with neutral pion trigger particles of transverse momenta $8 < p_{\mathrm{T}}^{\rm trig} < 16~\mathrm{GeV}/c$ and associated charged particles of $0.5 < p_{\mathrm{T}}^{\rm assoc} < 10~\mathrm{GeV}/c$ versus the azimuthal angle difference $\Delta\varphi$ at midrapidity in pp and central Pb-Pb collisions at $\sqrt{s_{\mathrm{NN}}}=2.76$ TeV with ALICE. 
The new measurements exploit associated charged hadrons down to $0.5~\mathrm{GeV}/c$, which significantly extends our previous measurement that only used charged hadrons above $3~\mathrm{GeV}/c$. 
After subtracting the contributions of the flow background, $v_2$ to $v_5$, the per-trigger yields are extracted for $|\Delta\varphi|<0.7$ on the near and for $|\Delta\varphi-\pi| < 1.1$ on the away side. 
The ratio of per-trigger yields in Pb--Pb to those in pp collisions, $I_{\mathrm{AA}}$, is measured on the near and away side for the $0$--$10$\% most central Pb--Pb collisions. 
On the away side, the per-trigger yields in Pb--Pb are strongly suppressed to the level of $I_{\mathrm{AA}} \approx 0.6$ for $p_{\mathrm{T}}^{\rm assoc} > 3~\mathrm{GeV}/c$, while with decreasing momenta an enhancement develops reaching about $5$ at low $p_{\mathrm{T}}^{\rm assoc}$. 
On the near side, an enhancement of $I_{\mathrm{AA}}$ between $1.2$ at the highest to $1.8$ at the lowest $p_{\mathrm{T}}^{\rm assoc}$ is observed. 
The data are compared to parton-energy-loss predictions of the JEWEL and AMPT event generators, as well as to a perturbative QCD calculation with medium-modified fragmentation functions.
All calculations qualitatively describe the away-side suppression at high $p_{\mathrm{T}}^{\rm assoc}$. 
Only AMPT captures the enhancement at low $p_{\mathrm{T}}^{\rm assoc}$, both on the near and away side.
However, it also underpredicts $I_{\mathrm{AA}}$ above $5$ GeV/$c$, in particular on the near-side.

%% file: pcontent.tex

\section{Introduction}
\label{sec:introduction}
Strongly interacting matter consisting of deconfined quarks and gluons, the quark-gluon plasma (QGP), is produced in high-energy heavy-ion~(HI) collisions at the Relativistic Heavy Ion Collider~(RHIC)~\cite{Adams:2005dq,Adcox:2004mh,Arsene:2004fa,Back:2004je} and at the Large Hadron Collider~(LHC)~\cite{Aamodt:2010jd,Chatrchyan:2011sx,Aad:2015wga,Aamodt:2010pa,ATLAS:2011ah,Chatrchyan:2012wg,ALICE:2011ab,Aad:2013xma,Chatrchyan:2013kba}. 
Among others, jet quenching~\cite{Gyulassy:1990ye,Wang:1991xy}, the phenomenon that high transverse momentum~($\pt$) partons suffer energy loss by medium-induced gluon radiation~\cite{Gyulassy:1993hr,Wang:1994fx} and collisions with medium constituents~\cite{Peshier:2006hi,Peigne:2008nd}, is widely considered as strong evidence for QGP formation.
Jet quenching has been observed at RHIC~\cite{Adcox:2001jp,Adler:2002tq,Adler:2002xw,Adcox:2002pe,Adler:2003qi,Adams:2003kv,Adams:2003im,Back:2003qr,Arsene:2003yk,Adler:2005ee,Adare:2006nr,Adams:2006yt,Adare:2007vu,Adare:2008ae,Adare:2008cg,Adare:2010ry,Adare:2012qi,STAR:2016jdz} and at the LHC~\cite{Aamodt:2010jd,Aad:2010bu,Chatrchyan:2011sx,Chatrchyan:2011pb,Aamodt:2011vg,CMS:2012aa,Chatrchyan:2012nia,Chatrchyan:2012gw,Chatrchyan:2012gt,Aad:2012vca,Chatrchyan:2013exa,Chatrchyan:2013kwa,Chatrchyan:2014ava,Aad:2014wha,Aad:2014bxa,Adam:2015ewa,Aad:2015wga} via measurements of inclusive hadron and jet production at high $\pt$, di-hadron angular correlations and di-jet energy imbalance, and via the modification of jet fragmentation functions.

In particular, measurements using two-particle angular correlations between trigger~(high-$\pt$) particles and associated particles have been extensively used to search for remnants of the radiated energy and the medium response to the high-$\pt$ parton.  
By varying the transverse momentum for trigger~($\ptt$) and associated~($\pta$) particles one can probe different momentum scales to study the interplay of soft and hard processes.
At RHIC, for a relatively low momentum range of $\ptt$ and $\pta$ below about $4~\gmom$, two-particle azimuthal angle correlations were found to be broadened and exhibiting a double-shoulder structure on the away side~\cite{Adler:2005ee,Adare:2007vu}.
These structures were originally described employing a variety of different mechanisms, like \v{C}erenkov gluon radiation~\cite{Koch:2005sx}, large angle gluon radiation~\cite{Vitev:2005yg,Polosa:2006hb}, Mach cone shockwave~\cite{CasalderreySolana:2006sq}, and jets deflected by the medium~\cite{Chiu:2006pu}. 
Later it was understood that azimuthal correlations spanning a long-range in pseudorapidity~($\eta$) are affected not only by the second~($v_2$) but also higher-order flow harmonics~($v_n$, $n\ge3$), which originate from anisotropic pressure gradients with respect to the initial-state symmetry planes~\cite{Alver:2010gr,Alver:2010dn}.
Taking into account these higher harmonics can account for most of the observed structures in the measured two-particle angular correlations.
Thus, possible jet-medium effects at low $\pt$ need to be studied after taking into account the anisotropic flow background including higher harmonics.

In this article, we present measurements of two-particle correlations with neutral pions~(\piz) of transverse momenta $8 < \ptt < 16~\gmom$ as trigger and charged hadrons of $0.5 < \pta < 10~\gmom$ as associated particles versus the azimuthal angle difference $\dphi$ at midrapidity in pp and central \PbPb\ collisions at $\snn=2.76$ TeV with ALICE~\cite{Aamodt:2008zz} at the LHC. 
The neutral pions are identified in the di-photon decay channel using a shower-shape and invariant-mass based identification technique of energy deposits reconstructed with the Electromagnetic Calorimeter (EMCal). 
The new measurement exploits associated hadrons reconstructed with the Inner Tracking System~(ITS) and Time Projection Chamber~(TPC) down to $0.5~\gmom$, and hence significantly extends our previous measurement~\cite{Aamodt:2011vg}, which only used charged hadrons above $3~\gmom$, to low $\pta$.
Furthermore, using $\piz$ as a reference avoids admixtures from changing particle composition of the trigger particle, and hence should simplify comparisons with calculations.
After subtracting the dominant background, induced by the anisotropic flow harmonics $v_{2}$ to $v_{5}$, the per-trigger yields are extracted for $|\dphi|<0.7$ on the near and for $|\dphi-\pi| < 1.1$ on the away side. 
The per-trigger yield modification factor, \Iaa, quantified as the ratio of per-trigger yields in \PbPb\ to those in pp collisions, is measured on the near and away side for the $0$--$10$\% most central \PbPb\ collisions.
The data are compared to parton-energy-loss model predictions using the JEWEL~\cite{Zapp:2012ak} and AMPT~\cite{Ma:2010dv} event generators, as well as to a perturbative QCD~(pQCD) calculation~\cite{Liu:2015vna} with medium-modified fragmentation functions.
Previously at RHIC, $\piz$-hadron correlations were also measured to study $\Iaa$ and jet fragmentation~\cite{Adare:2010ry,STAR:2016jdz}. 
Compared to these measurements, we lower the threshold for associated charged hadrons to $0.5~\gmom$ and substract the harmonic flow contributions up to the fifth order. 
Besides providing access to medium properties, measurements of $\piz$--hadron correlations determine the most important background contribution of direct photon--hadron correlation measurements\co{, which are considered as the ``golden channel'' to investigate medium properties, since the direct photon does not interact with the medium and therefore tags the kinematics of the initial hard scattering}~\cite{Adare:2012qi,STAR:2016jdz}.

The article is organized as follows.
\Section{sec:setup} briefly describes the experimental setup and data sets used. 
\Section{sec:analysis} discusses the neutral pion identification technique, the $\piz$--hadron correlation and $\Iaa$ measurements.
\Section{sec:results} presents the data and comparison with model calculations.
\Section{sec:summary} provides a summary.

\section{Experimental setup and datasets}
\label{sec:setup}
A detailed description of the ALICE detector systems and their performance can be found in~\cite{Aamodt:2008zz,Abelev:2014ffa}. 
The detectors used for the present analysis are briefly described here.
These are the ITS and the TPC for charged particle tracking, the EMCal for neutral pion reconstruction, and the forward scintillator arrays~(V0) and two Zero Degree Calorimeters~(ZDC) for online triggering as well as event selection and characterization. 

The tracking detectors are located inside a large solenoidal magnet providing a homogeneous field strength of $0.5$~T, and nominally provide reconstructed tracks within $|\eta|<0.9$ over the full azimuth. 
The ITS consists of six layers of silicon detectors.
The two inner layers are the Silicon Pixel Detector~(SPD), the two middle layers the Silicon Drift Detector~(SDD), and two outer layers the Silicon Strip Detector~(SSD).
The TPC provides tracking and particle identification by measuring the curvature of the tracks in the magnetic field and the specific energy loss ${\rm d}E/{\rm d}x$. 
The combined information of the ITS and TPC allows one to determine the momenta of charged particles in the region of $0.15$ to $100$ GeV/$c$ with a resolution of $1$ to $10$\%, respectively.
The EMCal is a Pb-scintillator sampling calorimeter used primarily to measure the energy deposit~(cluster) induced by electrons, positrons and photons.
It consists of $10$ active supermodules with a total of $11520$ individual cells, each covering an angular region of $\dphi\times\deta=0.014\times0.014$, and spans in total $100$ degrees in azimuth and $|\eta|<0.7$. 
Its energy resolution can be parameterized as $\frac{\sigma_E}{E} = \sqrt{A^2 + \frac{B^2}{E} +\frac{C^2}{E^2}}$\% with $A=1.68$, $B=11.27$ and $C=4.84$ for the deposited energy $E$ given in GeV~\cite{Abeysekara:2010ze}.
The V0 detectors, which are primarily used for triggering, event selection and event characterization, consist of two arrays of $32$ scintillator tiles each, covering the full azimuth within $2.8 < \eta < 5.1$~(V0-A) and $-3.7 < \eta < -1.7$~(V0-C).
In addition, two neutron ZDCs, located at $+114$~m~(ZNA) and $-114$~m~(ZNC) from the interaction point, are used for event selection in \PbPb\ collisions.  

The data used for the present analysis were collected during the 2011 LHC data taking periods with pp and \PbPb\ collisions at the centre-of-mass energy per nucleon--nucleon pair of $\snn=2.76$ TeV. 
In the case of pp collisions, the analyzed data were selected by the EMCal level-0 trigger requiring a single shower with an energy larger than $3.0$~GeV, in addition to the minimum bias trigger condition~(a hit in either V0-A, V0-C, or SPD). 
In the case of \PbPb\ collisions, the data were selected by an online trigger designed to select central collisions.
The trigger was selecting events based on the sum of amplitudes integrated in one LHC clock cycle~($25$~ns) online in the forward V0 detectors above a fixed threshold. 
Offline, when one can integrate the signal over several clock cycles, the trigger was found to be $100$\% efficient for $0$--$8$\%  and about $80$\% for $8$--$10$\% most central \PbPb\ collisions. 
The inefficiency in the 8--10\% range was estimated to lead to a negligible difference of less than $1$\% in the measured per-trigger yield.
For the offline analysis $0$--$10$\% central collisions were used as explained in detail in \Ref{Abelev:2013qoq}.
In both, the pp and \PbPb\ analyses, only events with a reconstructed vertex in $|z_{\rm vtx}|<10$~cm with respect to the nominal interaction vertex position along the beam direction were used. 
After all selection criteria, about 440K events in pp~(corresponding to $0.5$/nb) and 5.2M~(corresponding to $0.6/\mu$b) in \PbPb\ were kept for further analysis.

Neutral pions in $|\eta|<0.7$ are identified in the EMCal using the so called ``cluster splitting'' method, which aims to reconstruct a high $\pt$ $\piz$~(above $6~\gmom$) by first capturing both decay photons in a single, so called ``merged'' cluster, which then is split into two clusters, as further explained below.
Clusters are obtained by grouping all neighboring cells, whose calibrated energy is above $50$~($150$) MeV, starting from a seed cell with at least $100$~($300$) MeV for pp~(\PbPb) data.
A non-linearity correction, derived from electron test beam data, of about 7\% at $0.5$~GeV and negligible above $3$~GeV, is applied to the reconstructed cluster energy. 
Clusters from neutral particles are identified by requiring that the distance between the extrapolated track positions on the EMCal surface and the cluster fulfills the conditions $\deta>0.025$ and $\dphi>0.03$ for pp, and $\deta>0.03$ and $\dphi>0.035$ for \PbPb\ data.
Charged hadrons reconstructed with the ITS and TPC are selected by a hybrid approach designed to compensate local inefficiencies in the ITS. 
Two distinct track classes are accepted in the hybrid approach~\cite{Abelev:2014ffa}: 
      (i)~tracks containing at least three hits in the ITS, including at least one hit in the SPD, with momentum determined without the primary vertex constraint, 
 and (ii)~tracks containing less than three hits in the ITS or no hit in the SPD, with the primary vertex included in the momentum determination. 
Class~(i) contains 90\% and class~(ii) 10\% of all accepted tracks, independent of $\pt$. 
Track candidates are further required to have a Distance of Closest Approach~(DCA) to the primary vertex less than $2.4$~cm in the plane transverse to the beam, and less than $3.0$~cm in the beam direction. 
Accepted tracks are required to be in $|\eta|<0.8$ and $\pt > 0.5~\gmom$. 
Corrections for the detector response are obtained from Monte Carlo~(MC) detector simulations, reproducing the same conditions as during data taking.
In general, we use PYTHIA6~\cite{Sjostrand:2006za} for pp and HIJING~\cite{hijing} for \PbPb\ collisions as event generators, and GEANT3~\cite{geant3ref2} for particle transport through the detector. 

\section{Data analysis}
\label{sec:analysis}
Neutral pions are detected in the two photon decay channel $\piz\rightarrow\gamma\gamma$ measured in the EMCal using
\begin{equation}
 M_{\piz} = \sqrt{2E_{1}E_{2}(1-\cos\theta_{12})}\,,
 \label{eq:invariantmass}
\end{equation}
where $M_{\piz}$ is the reconstructed \piz\ mass, $E_{1}$ and $E_{2}$ are the measured energies of two photons, and $\theta_{12}$ is the opening angle between the photons measured in the laboratory frame. 
The opening angle decreases with increasing $\piz$ momentum due to the larger Lorentz boost. 
When the energy of the $\piz$ is larger than $5$--$6$ GeV, the decay photons are close enough that the electromagnetic showers they induce start to overlap in neighboring calorimeter cells of the EMCal.

Above $9$ GeV more than half of the $\piz$ deposit their energy in a single merged cluster.
Below $15$ GeV merged clusters from $\piz$ mostly have two local maxima~($\NLM=2$), while with increasing energy the showers further merge, leading to merged clusters from $\piz$ with mainly one local maximum~($\NLM=1$) above $25$~GeV.
Merged clusters can be identified based on their shower shape, characterized by the larger principal component squared of the cluster two-dimensional area in $\eta$ and $\phi$, $\lzt$~\cite{Alessandro:2006yt}.
To discriminate two-photon merged clusters from single-photon clusters, $\lzt$ is generally required to be greater than $0.3$. 
From detector simulations we deduced a tighter selection, requiring $\lztmin < \lzt < \lztmax$, where the minimum and maximum ranges are parameterized by $\exp\left(a+b\,E\right) + c + d\,E + e/E$ as a function of cluster energy~$E$~(in GeV).
For $\lztmin$, we use $a=2.135$, $b=-0.245$, $c=d=e=0$, while for $\lztmax$ the values depend on the number of local minima, and are $a=0.066$, $b=-0.020$, $c=-0.096$, $d=0.001$, and $e=9.91$ for $\NLM=1$, and $a=0.353$, $b=-0.0264$, $c=-0.524$, $d=0.006$, and $e=21.9$ for $\NLM=2$. 
Within $8<\pt<16~\gmom$, the range for neutral pions considered in this analysis, more than 80\% of the clusters have two local maxima. 

\begin{figure}[t!]
\begin{center}
\includegraphics[width=0.45\textwidth]{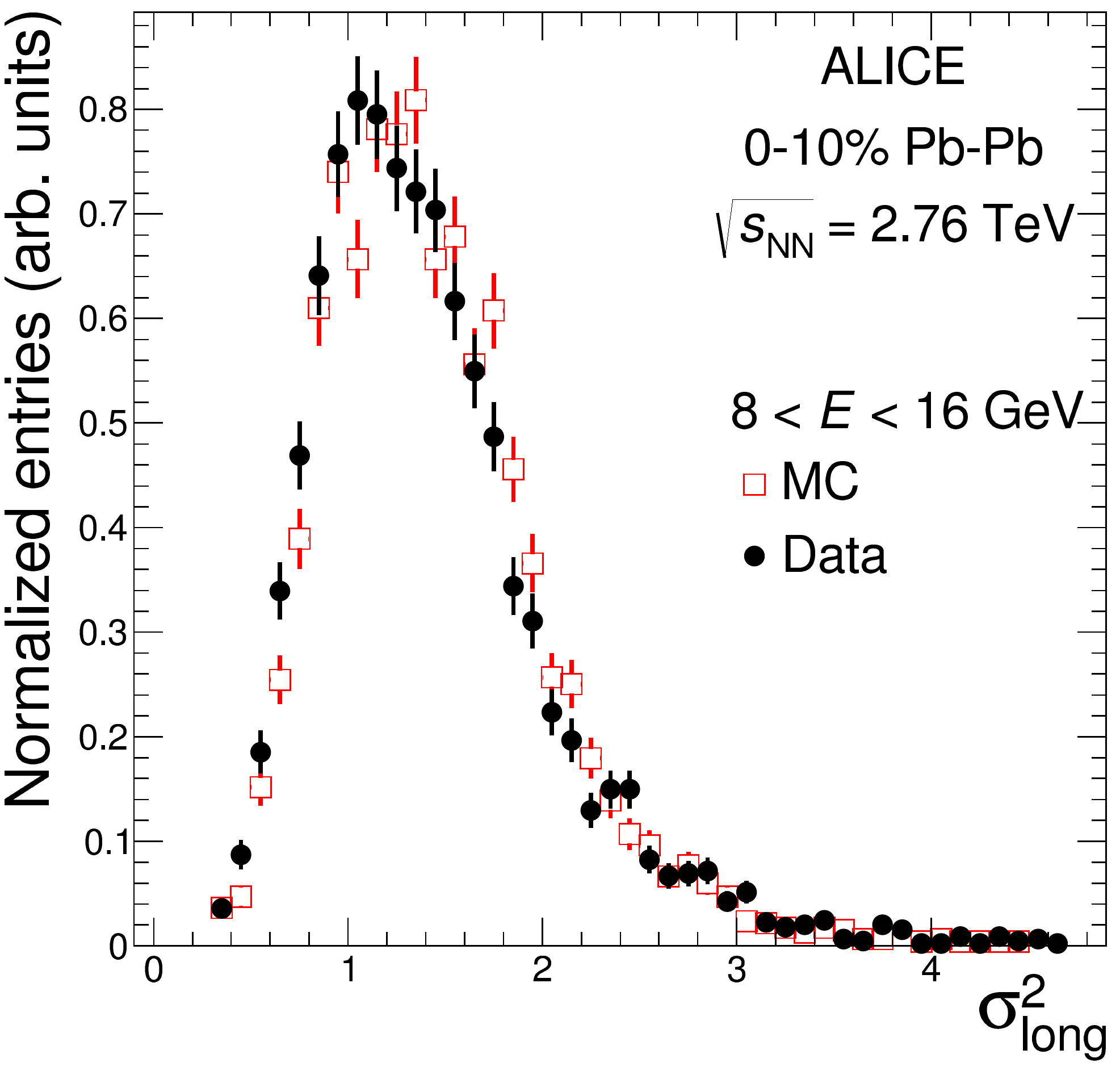}\hspace{0.5cm}
\includegraphics[width=0.45\textwidth]{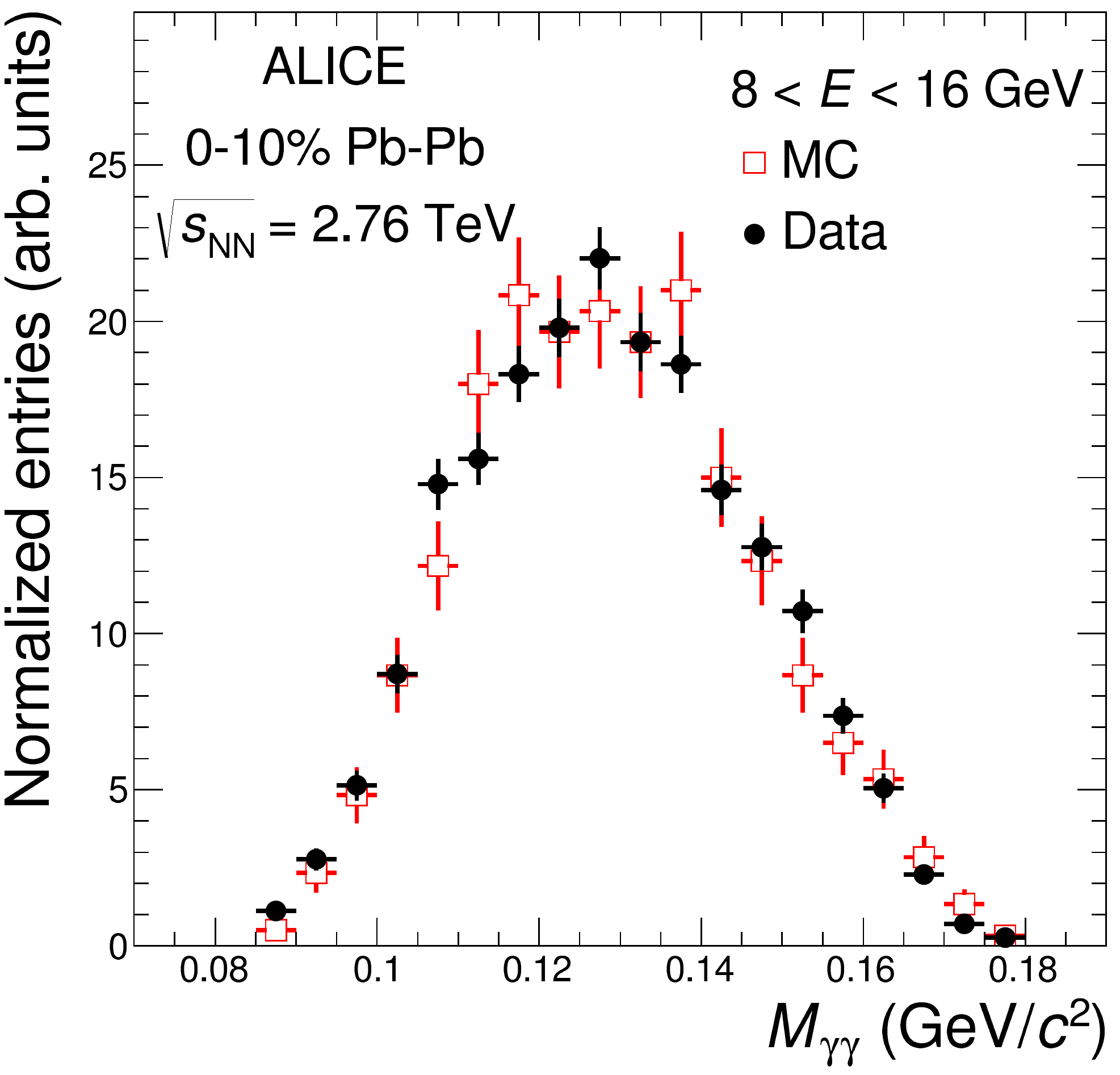}
\end{center}
\caption{Cluster shower shape~(left panel) and invariant mass~(right panel) distributions for $8<E<16$~GeV and $\NLM=2$ compared between reconstructed $\piz$ candidates in data and clusters originating from $\piz$ in HIJING for $0$--$10$\% \PbPb\ collisions. The distributions are shown after applying the energy-dependent selections on $\lzt$ and $M_{\gamma\gamma}$.} 
\label{fig:1}
\end{figure}

The merged cluster is subsequently split into two sub-clusters by grouping neighboring cells into $3\times3$ clusters centered around the two highest cells~(seeds) of the merged cluster.
Cells that are neighbor of both seeds are split based on the fraction of seed to cluster energy.
To select $\piz$ candidates, we use a $3\sigma$-wide window, $\av{M}-3\sigma < M_{\gamma\gamma} < \av{M}+3\sigma$, where the average~($\av{M}$) and the width~($\sigma$) of the mass distribution obtained from Gaussian fits depend on the energy of the cluster~(in GeV), and are each parameterized as $a + b \, E$.
The values for $a$ and $b$ are obtained from detector simulations for $\NLM=1$ and $2$, respectively, and are the same for pp and \PbPb\ data.
In the $\pt$ range relevant for the analysis, the parameters for $\av{M}$ are $a=0.044$ and $b=0.005$ for $\NLM=1$, and $a=0.115$, $b=0.001$ for $\NLM=2$, while for $\sigma$ they are $a=0.012$ and $b=0$ for $\NLM=1$, and $a=0.009$, $b=0.001$ for $\NLM=2$.
\Figure{fig:1} shows a comparison of $\lzt$ and $M_{\gamma\gamma}$ distributions for clusters with $8<E<16$~GeV and $\NLM=2$ between reconstructed $\piz$ candidates in data and clusters originating from $\piz$ in HIJING for $0$--$10$\% \PbPb\ collisions.
Since the invariant mass distribution is obtained by splitting individual clusters, there is no combinatorial background by construction. 
However, there is of course contamination in the signal region for example from decay photons, which needs to be estimated from Monte Carlo.

As commonly done~\cite{Oh:2016dxe}, the associated yield per trigger particle 
\begin{equation} \label{eq:yyield}
 Y(\dphi, \deta)=\frac{1}{\Ntrig}\frac{\dd^{2}\Nassoc}{\dd\dphi\dd\deta}=\frac{S(\deta,\dphi)}{M(\deta,\dphi)}
\end{equation}
is defined as the number of associated particles in intervals of azimuthal angle difference $\dphi=\varphi_{\rm trig}-\varphi_{\rm assoc}$ and pseudo-rapidity difference $\deta=\eta_{\rm trig}-\eta_{\rm assoc}$ relative to the number of trigger particles.
The trigger acceptance is $|\eta|<0.7$, while the associated particle acceptance is $|\eta|<0.8$.
The acceptance corrected yield can be obtained from the ratio of two-particle correlations of same $S$ and mixed events $M$.
The signal distribution $S(\deta,\dphi) = 1/\Ntrig \dd^2N_{\rm same}/\dd\deta\dd\dphi$ is the associated yield per trigger particle for particle pairs from the same event. 
The background distribution $M(\deta,\dphi) = \alpha\ \dd^2N_{\rm mixed}/\dd\deta\dd\dphi$ corrects for pair acceptance and pair efficiency.
It is constructed by correlating the trigger particles in one event with the associated particles from other events within similar multiplicity and $z$-vertex position intervals.
The factor $\alpha$ is chosen to normalize the background distribution such that it is unity for pairs where both particles go into approximately the same direction~(i.e.\ $\dphi\approx 0,\deta\approx 0$).
To account for different pair acceptance and pair efficiency as a function of $z_{\rm vtx}$, the yield is constructed for each $z_{\rm vtx}$ interval, and the final per-trigger yield is obtained by calculating the weighted average of the $z_{\rm vtx}$ intervals.
The final results are integrated over $\eta$ and provided as one-dimensional distribution, $C(\dphi)=\frac{1}{N_{\rm trig}}\frac{\dd\Nassoc}{\dd\dphi}$, for $8<\ptt<16~\gmom$ and various $\pta$ intervals between $0.5$ and $10~\gmom$.

Corrections for the detector response, which include $\piz$ reconstruction efficiency and purity, charged-particle tracking efficiency and contamination from secondary particles, as well as $\pt$ resolution are obtained from detector simulations.
The $\piz$ reconstruction efficiency, which is between $0.2$ and $0.3$ depending on $\pt$ and collision system, leads to only a small correction on the measured correlations of about $2$\%, since the per-trigger yield by definition is largely insensitive to the inefficiency of finding the trigger particle.
The $\piz$ purity, which in the momentum range of the measurement is about $90$\% in pp and $85$\% in \PbPb\ collisions, affects the measured correlations by $1$\%.
The $\pt$ resolution of reconstructed $\piz$ estimated from detector simulations is about 5\% and 10\% for pp and \PbPb\ collisions, respectively, slightly increasing with $\pt$.
The charged-particle tracking efficiency is about $75$--$85$\% depending on $\pt$ and collision system.
The contamination by secondary particles from particle-material interactions, conversions, and weak-decay products of long-lived particles is between $4$--$8$\%.
Both the tracking inefficiency and contamination, are corrected for in the measured correlations in intervals of $\pta$. 
The trigger- and associated-particle pair $\pt$ resolutions lead to a correction of less than $2.5$\%.

To obtain the jet-related contribution from the measured per-trigger yields, one usually subtracts non-jet related sources of particle production, 
\begin{equation} \label{eq:jyield}
 J(\dphi) = C(\dphi) - B(\dphi)\,,
\end{equation}
where $B(\dphi)$ denotes the background contribution. 
In pp collisions, typically a uniform background~($B_0$) originating from combinatorics is considered, and estimated employing the zero-yield-at-minimum~(ZYAM) method~\cite{Adler:2005ee}, i.e.\ essentially by estimating $B$ within $1<|\dphi|<\frac{\pi}{2}$.
In \PbPb\ collisions, in addition to a large combinatorial background, two-particle correlations are significantly affected by anisotropic flow~\cite{Aamodt:2011by}.
The anisotropic azimuthal correlations modulate the background according to  
\begin{equation} \label{eq:jbyield}
 B(\dphi) = B_{0}\,\left(1+2\sum_n\,V_n\,\cos(n\dphi)\right)\,,
\end{equation}
where $V_n\approx v^{\rm trig}_n \cdot v^{\rm assoc}_n$ is approximately given by the product of anisotropic flow coefficients for trigger and associated particles at their respective momenta.
In the subtraction, we take into account the most dominant contributions, $v_{2}$ to $v_{5}$, ignoring small deviations from factorization~\cite{Khachatryan:2015oea}. 
The data of $v_2$ for charged particles and for charged pions, which are used instead of the $v_2$ of $\piz$, are taken from~\Ref{Abelev:2012di}.
For $v_3$ to $v_5$ the data from \Ref{Aamodt:2011by} are used for both the neutral pions and charged particles.
The constant $B_{0}$ is determined by an average of three ways to obtain the ZYAM value, namely by
i)~a fit in $1<|\dphi|<\frac{\pi}{2}$, 
     ii)~smallest 8~(out of 60) values in full $\dphi$ range, 
and iii)~minima within $1<|\dphi|<\frac{\pi}{2}$ plus the two smallest points within $0.2$ around the minimum.
Finally, the jet-like correlation yields on the near and away side are estimated from \Eq{eq:jyield} by integrating  a region of $|\Delta\varphi| < 0.7$ and $|\Delta\varphi-\pi| < 1.1$, respectively. 
Modification of the jet-like pair yields can then be quantified as the ratio of the integrated jet-like yields in AA over pp, as
\begin{equation} \label{eq:iaa}
 \Iaa = \int_{X} J_{\rm AA}(\dphi){\rm d}\dphi / \int_{X} J_{\rm pp}(\dphi){\rm d}\dphi\,,  
\end{equation}
where $X$ denotes either the near-side~(NS) or the away-side~(AS) region.

\begin{figure}[t!f]
\centering
 \includegraphics[width=0.90\linewidth]{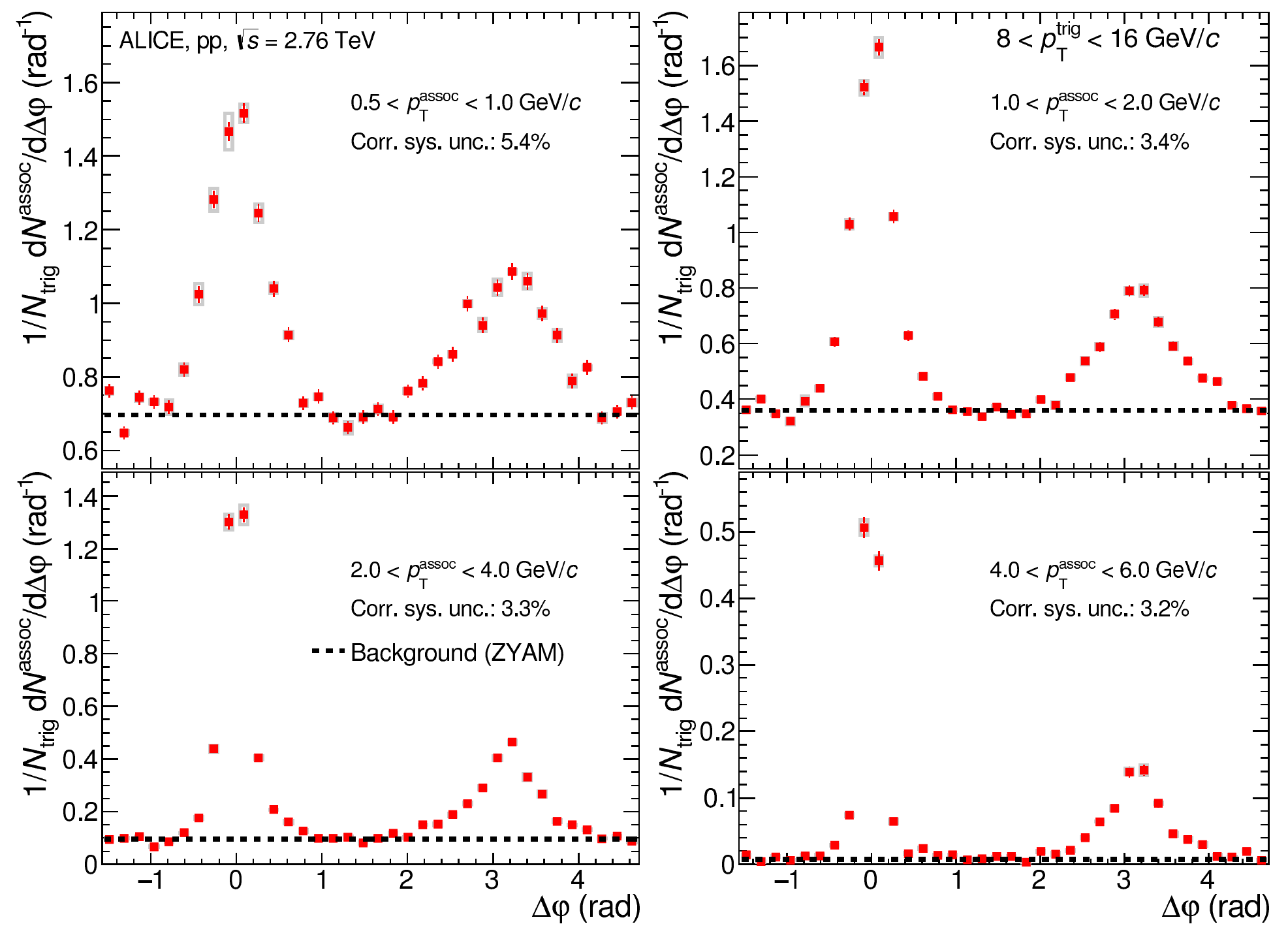}
 \caption{Charged-particle associated yields relative to $\pi^{0}$ trigger particles versus $\dphi$ in pp collisions at $\snn=2.76$ TeV. The $\pi^{0}$ trigger momentum range is $8< \ptt < 16~\gmom$, and associated charged particle ranges are $0.5 < \pta < 1$, $1 < \pta < 2$, $2 < \pta < 4$ and $4 < \pta < 6~\gmom$. The bars represent statistical uncertainties, the boxes uncorrelated systematic uncertainties. Dashed lines correspond to the estimated background using the ZYAM procedure described in the text. The range of the vertical axis is adjusted for each panel, and ``zero'' is not shown in all cases.
 \label{fig:dphipp}}
\end{figure} 

\begin{figure}[t!f]
\centering
 \includegraphics[width=0.90\linewidth]{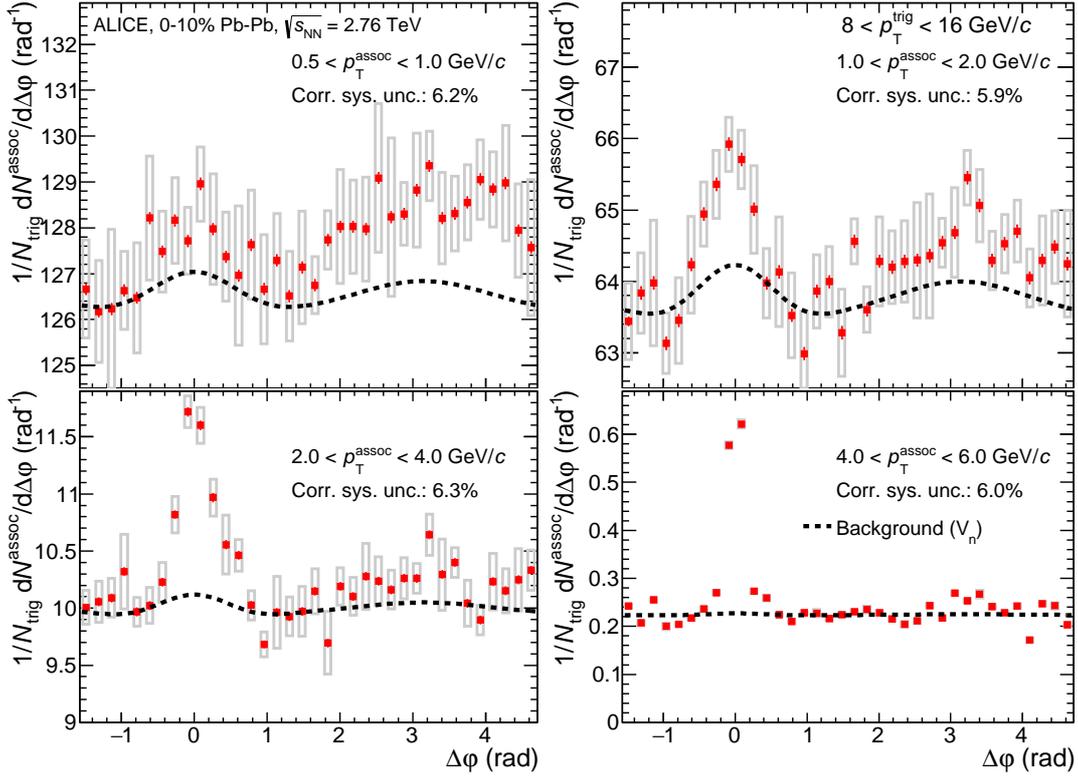}
 \caption{Charged-particle associated yields relative to $\pi^{0}$ trigger particles versus $\dphi$ in \mbox{$0$--$10$\%} most central \PbPb\ collisions at $\snn=2.76$ TeV. See caption of \Fig{fig:dphipp} for more information.}
 \label{fig:dphi}
\end{figure}
\begin{table}[t!]
\begin{center}
\begin{tabular}{l|c|c|c|c}
Source                &  $Y(\Delta\varphi)$ pp & $Y(\Delta\varphi)$ \PbPb\ & $\Iaa$~(NS) & $\Iaa$~(AS) \\
\hline
Tracking efficiency              &  $5.4$\%  &  $6.5$\%  &  $8.5$\%  &  $8.5$\%\\
MC closure                       &  $1.0$\%  &  $2.0$\%  &  $1.2$\%  &  $1.2$\%\\
TPC-only tracks                  &  $1.0$\%  &  $3.5$\%  &  $4.3$\%  &  $3.8$\%\\
Track contamination              &  $1.0$\%  &  $0.9$\%  &  $1.1$\%  &  $1.1$\%\\
Shower shape~($\lzt$)            &  $1.2$\%  &  $0.7$\%  &  $3.4$\%  &  $2.6$\%\\
Invariant mass window            &  $1.3$\%  &  $1.0$\%  &  $3.5$\%  &  $3.3$\%\\
Neutral pion  purity             &  $0.3$\%  &  $1.1$\%  &  $0.6$\%  &  $0.5$\%\\
Pair $\pt$ resolution            &  $1.0$\%  &  $1.1$\%  &  $0.3$\%  &  $0.3$\%\\
Pedestal determination           &  --       &  --       &  $9.4$\% &  $11.7$\%\\
Uncertainty on $v_n$             &  --       &  --       &  $7.1$\%  &  $5.1$\%\\
\hline
Total                            &  $6.7$\%  &  $7.4$\%  &  $12.6$\% &  $15.0$\% \\
\end{tabular}
\caption{\label{tab:syserrors}Summary of sources and assigned systematic uncertainties for the per-trigger yield in pp, and $0$--$10$\% \PbPb\ collisions, as well as \Iaa. For each source of systematic uncertainty and the total uncertainty listed, the maximum values of all $\pta$ intervals are given. Uncertainties on tracking efficiency and MC closure are correlated in $\dphi$. For $\Iaa$, pp and \PbPb\ yield uncertainties are assumed to be independent.}
\end{center}
\end{table} 

\section{Results}
\label{sec:results}
The per-trigger yields for neutral pion trigger particles with $8< \ptt < 16~\gmom$ and associated charged particles with $0.5 < \pta < 1$, $1 < \pta < 2$, $2 < \pta < 4$ and $4 < \pta < 6~\gmom$ are presented in \Fig{fig:dphipp} for pp and in \Fig{fig:dphi} for  $0$--$10$\% most central \PbPb\ collisions.
The estimated background from the ZYAM procedure is indicated by the dashed lines.
As explained in the previous section, a uniform background is considered in the case of pp, while for \PbPb\ data in addition the anisotropic flow contributions are taken into account.
Since the $v_n$ coefficients are small at high-$\ptt$ and $\pta$, a nearly flat background is observed for the $4 < \pta < 6~\gmom$ case, even in \PbPb\ collisions. 

Several sources of systematic uncertainty have been considered.
Since there is a $\pt$ dependence on the uncertainties, their maximum contribution to the per-trigger yields in pp and \PbPb\ collisions, as well as on the \Iaa\ further discussed below, are given in \Tab{tab:syserrors}.
The largest effect to the per-trigger yields arises from the uncertainty on the charged-particle tracking efficiency estimated from variations of the track selection and residual differences of MC closure tests. 
These uncertainties are correlated in $\dphi$, and their values~(added in quadrature) are explicitly reported in \Fig{fig:dphipp} and \Fig{fig:dphi}.
Uncertainties related to charged-particle tracking were further explored by repeating the full analysis with tracks reconstructed only by the TPC.  
Systematic uncertainties related to the $\piz$ identification were obtained by varying the criteria for $\lzt$ selection and the invariant mass window.
Uncertainties related to $\piz$ purity and $\pt$ resolution were assessed by varying the parameterizations, which were obtained from detector simulations and used for the respective corrections.
Total uncertainties were computed by adding the individual contributions in quadrature.
 
\begin{figure}[t]
\centering
 \includegraphics[width=0.48\linewidth]{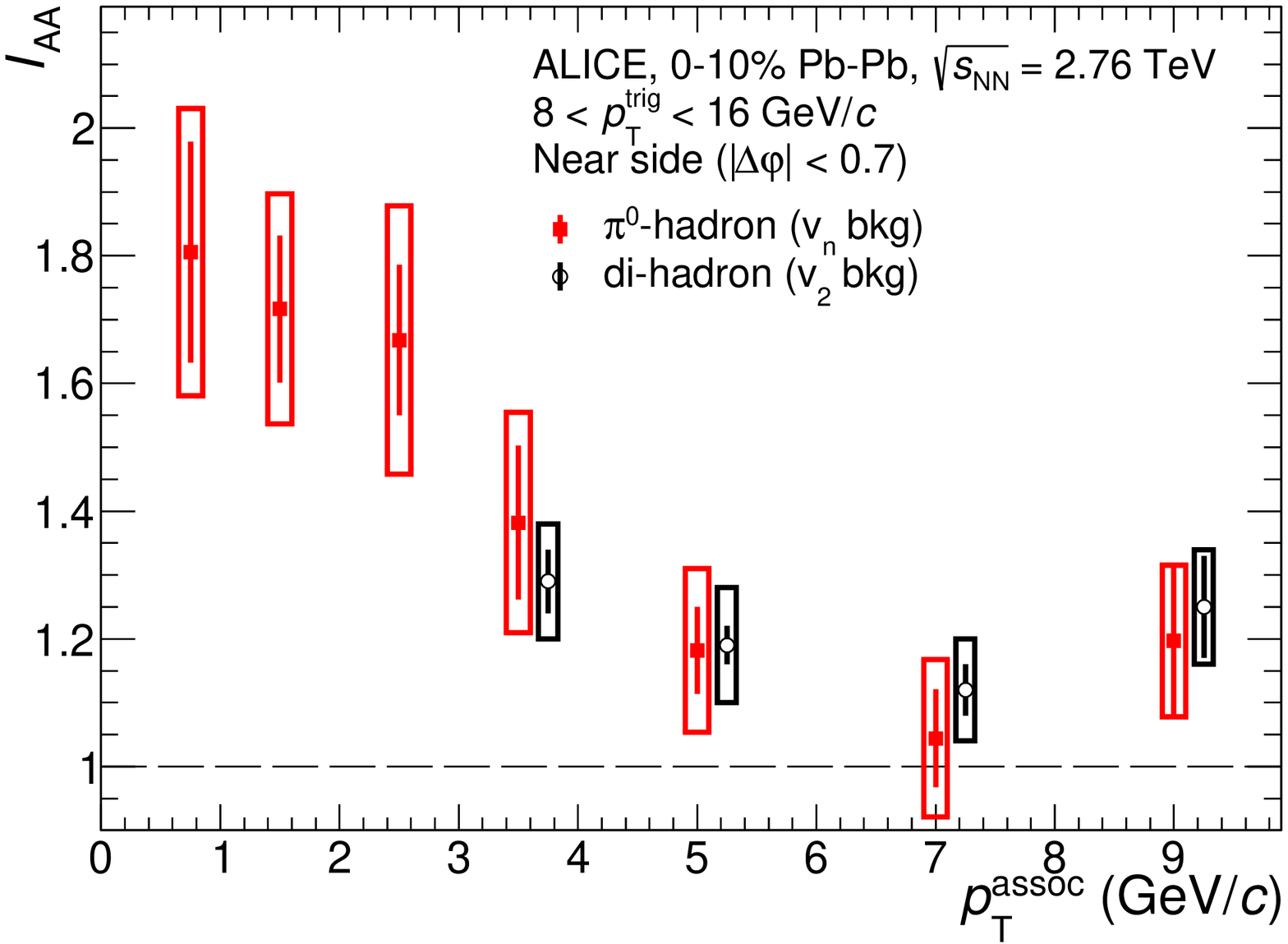}
 \includegraphics[width=0.48\linewidth]{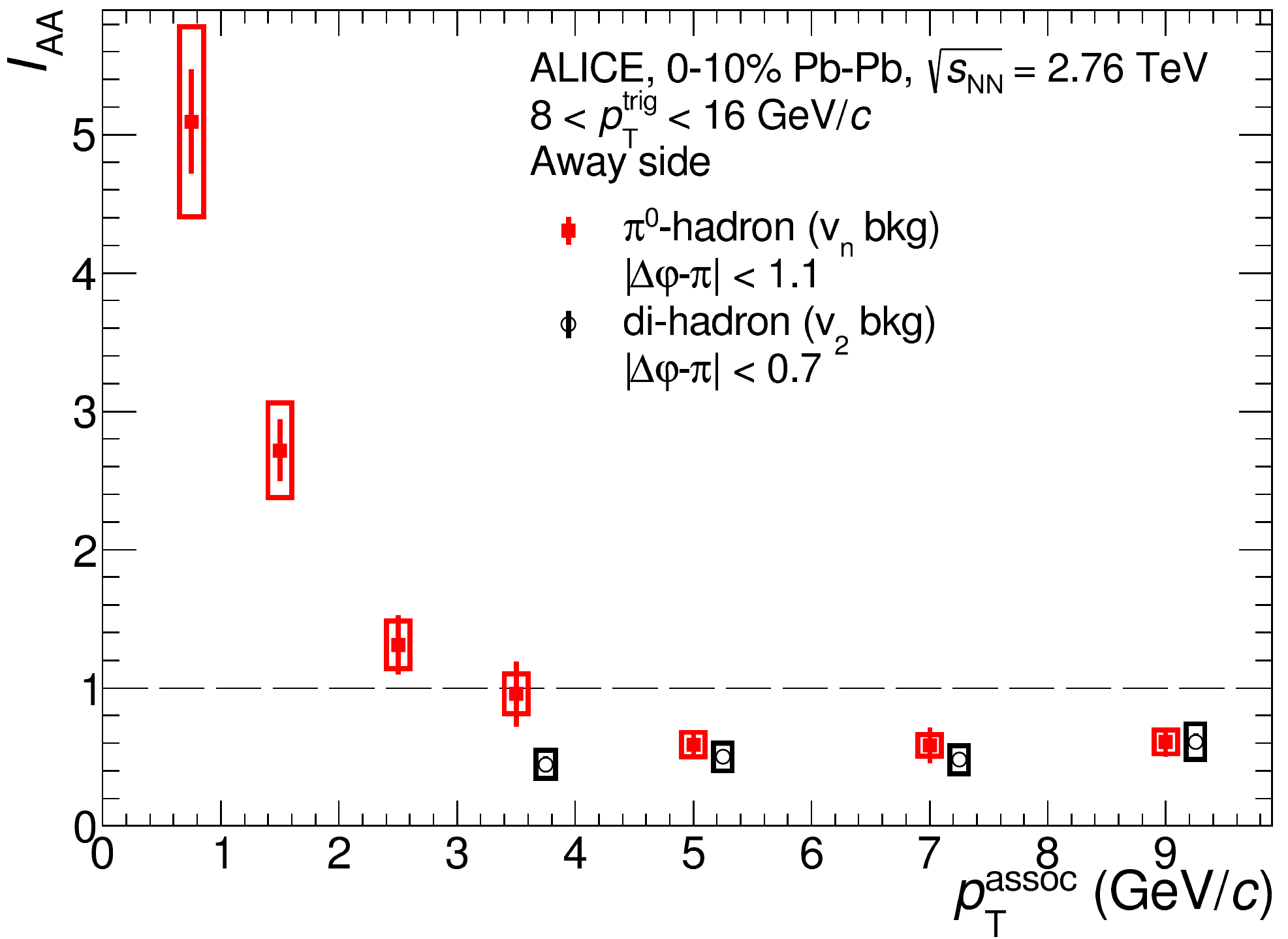}
 \caption{Per-trigger yield modification, $I_{\rm AA}$, on the near side~(left) and away side~(right) with trigger $\pi^{0}$ particle at $8 < \ptt < 16~\gmom$ for $0$--$10$\% \PbPb\ collisions at $\snn=2.76$ TeV. The data from our previous measurement using di-hadron correlations~\cite{Aamodt:2011vg} are slightly displaced for better visibility. The bars represent statistical and the boxes systematic uncertainties.
 \label{fig:IAA}}
\end{figure}

The modification of the per-trigger yield can be quantified as the ratio, $\Iaa$, of the integrated jet-like correlation yields in \PbPb\ over pp, as explained in the previous section~(see \Eq{eq:iaa}). 
\Figure{fig:IAA} presents the $\Iaa$ on the near side for $|\Delta\varphi| < 0.7$ and away side for $|\Delta\varphi-\pi| < 1.1$.
The uncertainty on $\Iaa$~(reported in \Tab{tab:syserrors}) is dominated by the uncertainty on the determination of $B_0$~(estimated from the difference of the 3 methods to extract the baseline) and the measured uncertainties on $v_n$, and hence it is largely uncorrelated across $\pta$. 
On the near side, the $\Iaa$ is found to be significantly larger than unity. 
The enhancement increases from $I_{\rm AA} \approx 1.2$ at high $\pta$ to $1.8$ at low $\pta$.
The data are consistent with our previous results extracted from di-hadron correlations above $3~\gmom$~\cite{Aamodt:2011vg}.
On the away side, $I_{\rm AA}$ is strongly enhanced below $3~\gmom$, reaching values up to $\Iaa\approx5$ at lowest $\pta$, while above $4~\gmom$ it is suppressed to about $0.6$.
As before, the data are compared to previous results using di-hadron correlations~\cite{Aamodt:2011vg}, which were obtained within a smaller integration region~($|\Delta\varphi| < 0.7$) and only taking into account $v_2$ in the ZYAM subtraction.
For $\pta>4~\gmom$, there is good agreement between the two sets of data\co{, as can be seen in the figure}, while for smaller $\pta$ the away-side peaks become wider and details of the ZYAM subtraction as well as the size of the integration region matter.
On the away side, the suppression at high $\pta$ is understood to originate from parton energy loss~\cite{Gyulassy:1990ye,Wang:1991xy,Gyulassy:1993hr,Wang:1994fx,Peshier:2006hi,Peigne:2008nd}, while the enhancement at low $\pta$ may involve an interplay of various contributions, such as $\kt$ broadening, medium-excitation, as well as fragments from radiated gluons~\cite{Wang:1998ww,Kopeliovich:2002yh,Vitev:2002pf,Vitev:2005yg,Ma:2010dv}.
The enhancement on the near side, first observed and discussed in \Ref{Aamodt:2011vg}, may also be related to the hot medium, inducing a change of the fragmentation function or the quark-to-gluon jet ratio. 

The observation of $\Iaa>1$ at low $\pt$ is consistent with the measured enhancement of low-$\pt$ particles from jet fragmentation in \PbPb\ relative to pp~\cite{Chatrchyan:2014ava,Aad:2014wha}.
At RHIC in \AuAu\ collisions at $200$~GeV for a similar range of $\ptt$ as used in the present measurement, $\Iaa$ on the away side was found to reach at most $2$--$3$~\cite{Adare:2010ry}, neglecting $v_3$ and higher orders harmonics in the background subtraction, while on the near side no significant enhancement was reported.

\begin{figure}[t!]
\centering
 \includegraphics[width=0.48\linewidth]{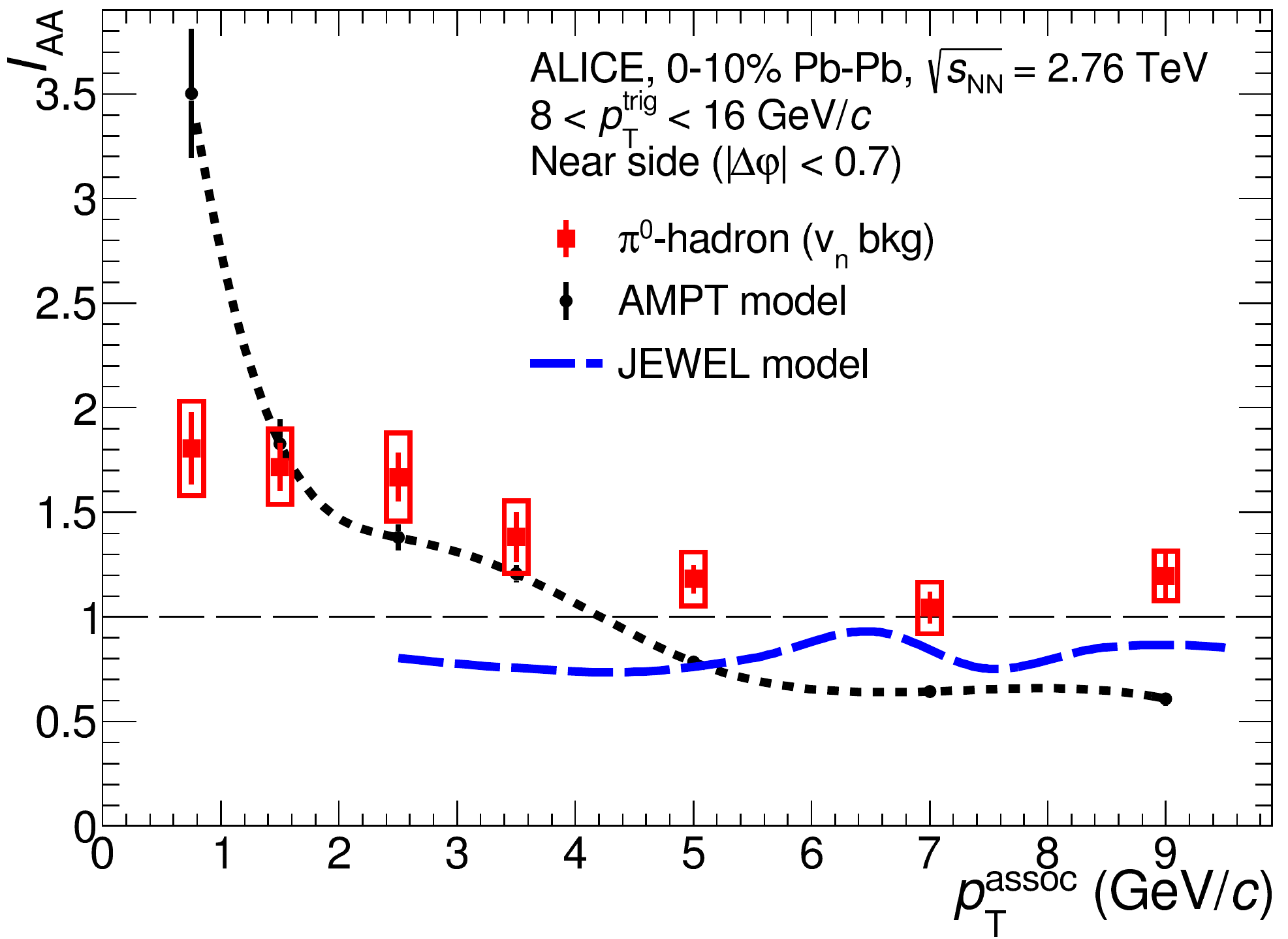}
 \includegraphics[width=0.48\linewidth]{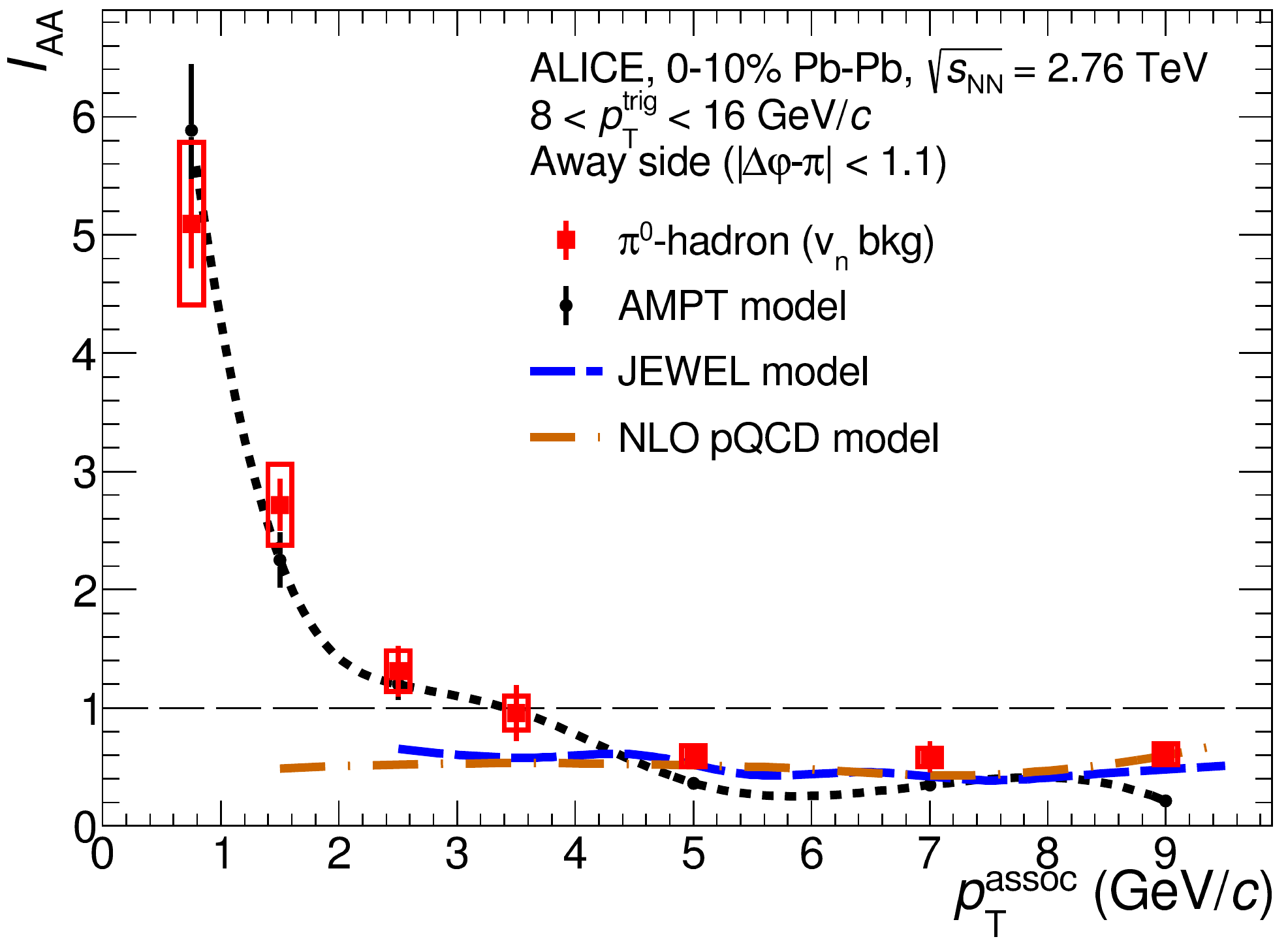}
 \caption{Per-trigger yield modification, $I_{\rm AA}$, on the near side~(left) and away side~(right) with trigger $\pi^{0}$ particle at $8 < \ptt < 16~\gmom$ for $0$--$10$\% \PbPb\ collisions at $\snn=2.76$ TeV. The data are compared to model calculations~\cite{Zapp:2012ak,Ma:2010dv,Liu:2015vna} as explained in the text. The bars represent and the boxes systematic uncertainties.
 \label{fig:IAAwM}} 
\end{figure}

In \Fig{fig:IAAwM} the data are compared to calculations using the JEWEL~\cite{Zapp:2012ak} and AMPT~\cite{Ma:2010dv} event generators, as well as pQCD calculation~\cite{Liu:2015vna}.
JEWEL~\cite{Zapp:2012ak} addresses the parton--medium interaction by giving a microscopic description of the transport coefficient, $\qhat$, which essentially defines the average energy loss per unit distance.
Hard scatters are generated according to Glauber collision geometry, and partons suffer from elastic and radiative energy loss in the medium, including a Monte Carlo implementation of LPM interference effects. 
The JEWEL calculation includes the so called ``recoil hadrons'', which are produced by fragmenting medium partons that interacted with the propagating hard parton.
AMPT~\cite{Lin:2004en} uses initial conditions of HIJING, followed by parton and hadron cascades with elastic scatterings for final-state interaction. 
String melting with a parton interaction cross section of $1.5$~mb and parton recombination for hadronization is used with parameters from \Ref{Xu:2011fi}\co{, which was shown to well describe collective flow in heavy-ion collisions at RHIC and LHC}\co{, and the flow background determined as in data following \Ref{Ma:2010dv}}. 
The pQCD calculation~\cite{Liu:2015vna} is performed at next-to-leading order~(NLO).
It uses nuclear parton distribution functions for initial-state cold nuclear matter effects, and a phenomenological model for medium-modified fragmentation functions\co{calculated in leading-order~(LO) at twist-4 in the high-twist approach of jet quenching}.
The evolution of bulk medium\co{ for parton propagation} is done with a 3+1 dimensional ideal hydrodynamic model, and the value $\qhat$ is consistent with that of the JET collaboration, which was extracted using experimental data~\cite{Burke:2013yra}.
The prediction for $\Iaa$ is only available for the away side, and done following \Ref{Zhang:2007ja}.

All calculations are able to qualitatively describe the suppression of $\Iaa$ at high $\pta$ on the away side, further corroborating the idea that the suppression is caused by parton energy loss in hot matter.
JEWEL and the pQCD calculation do not exhibit an increase at low $\pt$, while AMPT quantitatively describes the enhancement at the near (except at lowest $\pta$) and away side.
In AMPT the low-$\pta$ enhancement is attributed to the increase of soft particles as a result of the jet-medium interactions.
However, in particular on the near side for $\pta>5~\gmom$ AMPT predicts a strong suppression of $\Iaa$ down to about $0.6$, which clearly is not seen in the data.
Also on the away side AMPT tends to underpredict the \Iaa for $\pta>5~\gmom$. 
Both defects, which may be related to the fact that AMPT was found to overpredict the single-particle suppression in central \PbPb\ collisions~\cite{Pal:2012gf}, indicate that the description implemented in AMPT is not complete.

\section{Summary}
\label{sec:summary}
Two-particle correlations with neutral pions of transverse momenta $8 < \ptt < 16~\gmom$ as trigger and charged hadrons of $0.5 < \pta < 10~\gmom$ as associated particles versus azimuthal angle difference $\dphi$ at midrapidity in pp~(\Fig{fig:dphipp}) and central \PbPb~(\Fig{fig:dphi}) collisions at $\snn=2.76$ TeV have been measured.
The per-trigger yields have been extracted for $|\dphi|<0.7$ on the near and for $|\dphi-\pi| < 1.1$ on the away side, after subtracting the contributions of the flow harmonics, $v_{2}$ up to $v_{5}$~(\Fig{fig:dphi}). 
The per-trigger yield modification factor, $\Iaa$, quantified as the ratio of per-trigger yields in \PbPb\ to that in pp collisions, has been measured for the near and away side in $0$--$10$\% most central \PbPb\ collisions~(\Fig{fig:IAA}). 
On the away side, the per-trigger yields in \PbPb\ are strongly suppressed to the level of $\Iaa \approx 0.6$ for $\pta > 3~\gmom$, while with decreasing momenta an enhancement develops reaching about $5.2$ at lowest \pta.
On the near side, an enhancement of $\Iaa$ between $1.2$ to $1.8$ at lowest $\pta$ is observed.
The data are compared to predictions of the JEWEL and AMPT event generators, as well as a pQCD calculation at next-to-leading order with medium-modified fragmentation functions~(\Fig{fig:IAAwM}).
All calculations are able to qualitatively describe the away-side suppression at high $\pta$. 
Only AMPT is able to capture the enhancement at low $\pta$, both on near and away side.
However, it also underpredicts $\Iaa$ above $5$ GeV/$c$, in particular on the near-side.
The coincidence of the away-side suppression at high $\pt$ and the large enhancement at low $\pt$ on the near and away side is suggestive of a common underlying mechanism, likely related to the energy lost by high momentum partons.
The data hence provide a good testing ground to constrain model calculations which aim to fully describe jet--medium interactions.

%% file: acknowledgements.tex

The ALICE Collaboration would like to thank all its engineers and technicians for their invaluable contributions to the construction of the experiment and the CERN accelerator teams for the outstanding performance of the LHC complex.
The ALICE Collaboration gratefully acknowledges the resources and support provided by all Grid centres and the Worldwide LHC Computing Grid (WLCG) collaboration.
The ALICE Collaboration acknowledges the following funding agencies for their support in building and running the ALICE detector:
A. I. Alikhanyan National Science Laboratory (Yerevan Physics Institute) Foundation (ANSL), State Committee of Science and World Federation of Scientists (WFS), Armenia;
Austrian Academy of Sciences and Nationalstiftung f\"{u}r Forschung, Technologie und Entwicklung, Austria;
Conselho Nacional de Desenvolvimento Cient\'{\i}fico e Tecnol\'{o}gico (CNPq), Financiadora de Estudos e Projetos (Finep) and Funda\c{c}\~{a}o de Amparo \`{a} Pesquisa do Estado de S\~{a}o Paulo (FAPESP), Brazil;
Ministry of Education of China (MOE of China), Ministry of Science \& Technology of China (MOST of China) and National Natural Science Foundation of China (NSFC), China;
Ministry of Science, Education and Sport and Croatian Science Foundation, Croatia;
Centro de Investigaciones Energ\'{e}ticas, Medioambientales y Tecnol\'{o}gicas (CIEMAT), Cuba;
Ministry of Education, Youth and Sports of the Czech Republic, Czech Republic;
Danish National Research Foundation (DNRF), The Carlsberg Foundation and The Danish Council for Independent Research | Natural Sciences, Denmark;
Helsinki Institute of Physics (HIP), Finland;
Commissariat \`{a} l'Energie Atomique (CEA) and Institut National de Physique Nucl\'{e}aire et de Physique des Particules (IN2P3) and Centre National de la Recherche Scientifique (CNRS), France;
Bundesministerium f\"{u}r Bildung, Wissenschaft, Forschung und Technologie (BMBF) and GSI Helmholtzzentrum f\"{u}r Schwerionenforschung GmbH, Germany;
Ministry of Education, Research and Religious Affairs, Greece;
National Research, Development and Innovation Office, Hungary;
Department of Atomic Energy Government of India (DAE), India;
Indonesian Institute of Science, Indonesia;
Centro Fermi - Museo Storico della Fisica e Centro Studi e Ricerche Enrico Fermi and Istituto Nazionale di Fisica Nucleare (INFN), Italy;
Institute for Innovative Science and Technology , Nagasaki Institute of Applied Science (IIST), Japan Society for the Promotion of Science (JSPS) KAKENHI and Japanese Ministry of Education, Culture, Sports, Science and Technology (MEXT), Japan;
Consejo Nacional de Ciencia (CONACYT) y Tecnolog\'{i}a, through Fondo de Cooperaci\'{o}n Internacional en Ciencia y Tecnolog\'{i}a (FONCICYT) and Direcci\'{o}n General de Asuntos del Personal Academico (DGAPA), Mexico;
Nationaal instituut voor subatomaire fysica (Nikhef), Netherlands;
The Research Council of Norway, Norway;
Commission on Science and Technology for Sustainable Development in the South (COMSATS), Pakistan;
Pontificia Universidad Cat\'{o}lica del Per\'{u}, Peru;
Ministry of Science and Higher Education and National Science Centre, Poland;
Ministry of Education and Scientific Research, Institute of Atomic Physics and Romanian National Agency for Science, Technology and Innovation, Romania;
Joint Institute for Nuclear Research (JINR), Ministry of Education and Science of the Russian Federation and National Research Centre Kurchatov Institute, Russia;
Ministry of Education, Science, Research and Sport of the Slovak Republic, Slovakia;
National Research Foundation of South Africa, South Africa;
Korea Institute of Science and Technology Information and National Research Foundation of Korea (NRF), South Korea;
Centro de Investigaciones Energ\'{e}ticas, Medioambientales y Tecnol\'{o}gicas (CIEMAT) and Ministerio de Ciencia e Innovacion, Spain;
Knut \& Alice Wallenberg Foundation (KAW) and Swedish Research Council (VR), Sweden;
European Organization for Nuclear Research, Switzerland;
National Science and Technology Development Agency (NSDTA), Office of the Higher Education Commission under NRU project of Thailand and Suranaree University of Technology (SUT), Thailand;
Turkish Atomic Energy Agency (TAEK), Turkey;
National Academy of  Sciences of Ukraine, Ukraine;
Science and Technology Facilities Council (STFC), United Kingdom;
National Science Foundation of the United States of America (NSF) and United States Department of Energy, Office of Nuclear Physics (DOE NP), United States.

%% file: Alice_Authorlist_2016-Aug-01.tex


\begingroup
\small
\begin{flushleft}
J.~Adam$^\textrm{\scriptsize 39}$\textsuperscript{,}$^\textrm{\scriptsize 88}$,
D.~Adamov\'{a}$^\textrm{\scriptsize 85}$,
M.M.~Aggarwal$^\textrm{\scriptsize 89}$,
G.~Aglieri Rinella$^\textrm{\scriptsize 35}$,
M.~Agnello$^\textrm{\scriptsize 31}$\textsuperscript{,}$^\textrm{\scriptsize 112}$,
N.~Agrawal$^\textrm{\scriptsize 48}$,
Z.~Ahammed$^\textrm{\scriptsize 136}$,
S.~Ahmad$^\textrm{\scriptsize 18}$,
S.U.~Ahn$^\textrm{\scriptsize 69}$,
S.~Aiola$^\textrm{\scriptsize 140}$
% ...
\renewcommand\labelenumi{\textsuperscript{\theenumi}~}

\section*{Affiliation notes}
\renewcommand\theenumi{\roman{enumi}}
\begin{Authlist}
\item \Adef{0}Deceased
\item \Adef{idp1839312}{Also at: Georgia State University, Atlanta, Georgia, United States}
\item \Adef{idp3273760}{Also at: Also at Department of Applied Physics, Aligarh Muslim University, Aligarh, India}
\item \Adef{idp4023728}{Also at: M.V. Lomonosov Moscow State University, D.V. Skobeltsyn Institute of Nuclear, Physics, Moscow, Russia}
\end{Authlist}

\section*{Collaboration Institutes}
\renewcommand\theenumi{\arabic{enumi}~}

$^{1}$A.I. Alikhanyan National Science Laboratory (Yerevan Physics Institute) Foundation, Yerevan, Armenia
\\
$^{2}$Benem\'{e}rita Universidad Aut\'{o}noma de Puebla, Puebla, Mexico
\\
$^{3}$Bogolyubov Institute for Theoretical Physics, Kiev, Ukraine
\\
$^{4}$Bose Institute, Department of Physics 
and Centre for Astroparticle Physics and Space Science (CAPSS), Kolkata, India
\\
$^{5}$Budker Institute for Nuclear Physics, Novosibirsk, Russia
\\
$^{6}$California Polytechnic State University, San Luis Obispo, California, United States
\\
$^{7}$Central China Normal University, Wuhan, China
\\
$^{8}$Centre de Calcul de l'IN2P3, Villeurbanne, Lyon, France
\\
$^{9}$Centro de Aplicaciones Tecnol\'{o}gicas y Desarrollo Nuclear (CEADEN), Havana, Cuba
\\
$^{10}$Centro de Investigaciones Energ\'{e}ticas Medioambientales y Tecnol\'{o}gicas (CIEMAT), Madrid, Spain
\\
$^{11}$Centro de Investigaci\'{o}n y de Estudios Avanzados (CINVESTAV), Mexico City and M\'{e}rida, Mexico
\\
$^{12}$Centro Fermi - Museo Storico della Fisica e Centro Studi e Ricerche ``Enrico Fermi', Rome, Italy
\\
$^{13}$Chicago State University, Chicago, Illinois, United States
\\
$^{14}$China Institute of Atomic Energy, Beijing, China
\\
$^{15}$Commissariat \`{a} l'Energie Atomique, IRFU, Saclay, France
\\
$^{16}$COMSATS Institute of Information Technology (CIIT), Islamabad, Pakistan
\\
$^{17}$Departamento de F\'{\i}sica de Part\'{\i}culas and IGFAE, Universidad de Santiago de Compostela, Santiago de Compostela, Spain
\\
$^{18}$Department of Physics, Aligarh Muslim University, Aligarh, India
\\
$^{19}$Department of Physics, Ohio State University, Columbus, Ohio, United States
\\
$^{20}$Department of Physics, Sejong University, Seoul, South Korea
\\
$^{21}$Department of Physics, University of Oslo, Oslo, Norway
\\
$^{22}$Department of Physics and Technology, University of Bergen, Bergen, Norway
\\
$^{23}$Dipartimento di Fisica dell'Universit\`{a} 'La Sapienza'
and Sezione INFN, Rome, Italy
\\
$^{24}$Dipartimento di Fisica dell'Universit\`{a}
and Sezione INFN, Cagliari, Italy
\\
$^{25}$Dipartimento di Fisica dell'Universit\`{a}
and Sezione INFN, Trieste, Italy
\\
$^{26}$Dipartimento di Fisica dell'Universit\`{a}
and Sezione INFN, Turin, Italy
\\
$^{27}$Dipartimento di Fisica e Astronomia dell'Universit\`{a}
and Sezione INFN, Bologna, Italy
\\
$^{28}$Dipartimento di Fisica e Astronomia dell'Universit\`{a}
and Sezione INFN, Catania, Italy
\\
$^{29}$Dipartimento di Fisica e Astronomia dell'Universit\`{a}
and Sezione INFN, Padova, Italy
\\
$^{30}$Dipartimento di Fisica `E.R.~Caianiello' dell'Universit\`{a}
and Gruppo Collegato INFN, Salerno, Italy
\\
$^{31}$Dipartimento DISAT del Politecnico and Sezione INFN, Turin, Italy
\\
$^{32}$Dipartimento di Scienze e Innovazione Tecnologica dell'Universit\`{a} del Piemonte Orientale and INFN Sezione di Torino, Alessandria, Italy
\\
$^{33}$Dipartimento Interateneo di Fisica `M.~Merlin'
and Sezione INFN, Bari, Italy
\\
$^{34}$Division of Experimental High Energy Physics, University of Lund, Lund, Sweden
\\
$^{35}$European Organization for Nuclear Research (CERN), Geneva, Switzerland
\\
$^{36}$Excellence Cluster Universe, Technische Universit\"{a}t M\"{u}nchen, Munich, Germany
\\
$^{37}$Faculty of Engineering, Bergen University College, Bergen, Norway
\\
$^{38}$Faculty of Mathematics, Physics and Informatics, Comenius University, Bratislava, Slovakia
\\
$^{39}$Faculty of Nuclear Sciences and Physical Engineering, Czech Technical University in Prague, Prague, Czech Republic
\\
$^{40}$Faculty of Science, P.J.~\v{S}af\'{a}rik University, Ko\v{s}ice, Slovakia
\\
$^{41}$Faculty of Technology, Buskerud and Vestfold University College, Tonsberg, Norway
\\
$^{42}$Frankfurt Institute for Advanced Studies, Johann Wolfgang Goethe-Universit\"{a}t Frankfurt, Frankfurt, Germany
\\
$^{43}$Gangneung-Wonju National University, Gangneung, South Korea
\\
$^{44}$Gauhati University, Department of Physics, Guwahati, India
\\
$^{45}$Helmholtz-Institut f\"{u}r Strahlen- und Kernphysik, Rheinische Friedrich-Wilhelms-Universit\"{a}t Bonn, Bonn, Germany
\\
$^{46}$Helsinki Institute of Physics (HIP), Helsinki, Finland
\\
$^{47}$Hiroshima University, Hiroshima, Japan
\\
$^{48}$Indian Institute of Technology Bombay (IIT), Mumbai, India
\\
$^{49}$Indian Institute of Technology Indore, Indore, India
\\
$^{50}$Indonesian Institute of Sciences, Jakarta, Indonesia
\\
$^{51}$Inha University, Incheon, South Korea
\\
$^{52}$Institut de Physique Nucl\'eaire d'Orsay (IPNO), Universit\'e Paris-Sud, CNRS-IN2P3, Orsay, France
\\
$^{53}$Institute for Nuclear Research, Academy of Sciences, Moscow, Russia
\\
$^{54}$Institute for Subatomic Physics of Utrecht University, Utrecht, Netherlands
\\
$^{55}$Institute for Theoretical and Experimental Physics, Moscow, Russia
\\
$^{56}$Institute of Experimental Physics, Slovak Academy of Sciences, Ko\v{s}ice, Slovakia
\\
$^{57}$Institute of Physics, Academy of Sciences of the Czech Republic, Prague, Czech Republic
\\
$^{58}$Institute of Physics, Bhubaneswar, India
\\
$^{59}$Institute of Space Science (ISS), Bucharest, Romania
\\
$^{60}$Institut f\"{u}r Informatik, Johann Wolfgang Goethe-Universit\"{a}t Frankfurt, Frankfurt, Germany
\\
$^{61}$Institut f\"{u}r Kernphysik, Johann Wolfgang Goethe-Universit\"{a}t Frankfurt, Frankfurt, Germany
\\
$^{62}$Institut f\"{u}r Kernphysik, Westf\"{a}lische Wilhelms-Universit\"{a}t M\"{u}nster, M\"{u}nster, Germany
\\
$^{63}$Instituto de Ciencias Nucleares, Universidad Nacional Aut\'{o}noma de M\'{e}xico, Mexico City, Mexico
\\
$^{64}$Instituto de F\'{\i}sica, Universidad Nacional Aut\'{o}noma de M\'{e}xico, Mexico City, Mexico
\\
$^{65}$Institut Pluridisciplinaire Hubert Curien (IPHC), Universit\'{e} de Strasbourg, CNRS-IN2P3, Strasbourg, France
\\
$^{66}$iThemba LABS, National Research Foundation, Somerset West, South Africa
\\
$^{67}$Joint Institute for Nuclear Research (JINR), Dubna, Russia
\\
$^{68}$Konkuk University, Seoul, South Korea
\\
$^{69}$Korea Institute of Science and Technology Information, Daejeon, South Korea
\\
$^{70}$KTO Karatay University, Konya, Turkey
\\
$^{71}$Laboratoire de Physique Corpusculaire (LPC), Clermont Universit\'{e}, Universit\'{e} Blaise Pascal, CNRS--IN2P3, Clermont-Ferrand, France
\\
$^{72}$Laboratoire de Physique Subatomique et de Cosmologie, Universit\'{e} Grenoble-Alpes, CNRS-IN2P3, Grenoble, France
\\
$^{73}$Laboratori Nazionali di Frascati, INFN, Frascati, Italy
\\
$^{74}$Laboratori Nazionali di Legnaro, INFN, Legnaro, Italy
\\
$^{75}$Lawrence Berkeley National Laboratory, Berkeley, California, United States
\\
$^{76}$Moscow Engineering Physics Institute, Moscow, Russia
\\
$^{77}$Nagasaki Institute of Applied Science, Nagasaki, Japan
\\
$^{78}$National Centre for Nuclear Studies, Warsaw, Poland
\\
$^{79}$National Institute for Physics and Nuclear Engineering, Bucharest, Romania
\\
$^{80}$National Institute of Science Education and Research, Bhubaneswar, India
\\
$^{81}$National Research Centre Kurchatov Institute, Moscow, Russia
\\
$^{82}$Niels Bohr Institute, University of Copenhagen, Copenhagen, Denmark
\\
$^{83}$Nikhef, Nationaal instituut voor subatomaire fysica, Amsterdam, Netherlands
\\
$^{84}$Nuclear Physics Group, STFC Daresbury Laboratory, Daresbury, United Kingdom
\\
$^{85}$Nuclear Physics Institute, Academy of Sciences of the Czech Republic, \v{R}e\v{z} u Prahy, Czech Republic
\\
$^{86}$Oak Ridge National Laboratory, Oak Ridge, Tennessee, United States
\\
$^{87}$Petersburg Nuclear Physics Institute, Gatchina, Russia
\\
$^{88}$Physics Department, Creighton University, Omaha, Nebraska, United States
\\
$^{89}$Physics Department, Panjab University, Chandigarh, India
\\
$^{90}$Physics Department, University of Athens, Athens, Greece
\\
$^{91}$Physics Department, University of Cape Town, Cape Town, South Africa
\\
$^{92}$Physics Department, University of Jammu, Jammu, India
\\
$^{93}$Physics Department, University of Rajasthan, Jaipur, India
\\
$^{94}$Physikalisches Institut, Eberhard Karls Universit\"{a}t T\"{u}bingen, T\"{u}bingen, Germany
\\
$^{95}$Physikalisches Institut, Ruprecht-Karls-Universit\"{a}t Heidelberg, Heidelberg, Germany
\\
$^{96}$Physik Department, Technische Universit\"{a}t M\"{u}nchen, Munich, Germany
\\
$^{97}$Purdue University, West Lafayette, Indiana, United States
\\
$^{98}$Pusan National University, Pusan, South Korea
\\
$^{99}$Research Division and ExtreMe Matter Institute EMMI, GSI Helmholtzzentrum f\"ur Schwerionenforschung, Darmstadt, Germany
\\
$^{100}$Rudjer Bo\v{s}kovi\'{c} Institute, Zagreb, Croatia
\\
$^{101}$Russian Federal Nuclear Center (VNIIEF), Sarov, Russia
\\
$^{102}$Saha Institute of Nuclear Physics, Kolkata, India
\\
$^{103}$School of Physics and Astronomy, University of Birmingham, Birmingham, United Kingdom
\\
$^{104}$Secci\'{o}n F\'{\i}sica, Departamento de Ciencias, Pontificia Universidad Cat\'{o}lica del Per\'{u}, Lima, Peru
\\
$^{105}$Sezione INFN, Bari, Italy
\\
$^{106}$Sezione INFN, Bologna, Italy
\\
$^{107}$Sezione INFN, Cagliari, Italy
\\
$^{108}$Sezione INFN, Catania, Italy
\\
$^{109}$Sezione INFN, Padova, Italy
\\
$^{110}$Sezione INFN, Rome, Italy
\\
$^{111}$Sezione INFN, Trieste, Italy
\\
$^{112}$Sezione INFN, Turin, Italy
\\
$^{113}$SSC IHEP of NRC Kurchatov institute, Protvino, Russia
\\
$^{114}$Stefan Meyer Institut f\"{u}r Subatomare Physik (SMI), Vienna, Austria
\\
$^{115}$SUBATECH, Ecole des Mines de Nantes, Universit\'{e} de Nantes, CNRS-IN2P3, Nantes, France
\\
$^{116}$Suranaree University of Technology, Nakhon Ratchasima, Thailand
\\
$^{117}$Technical University of Ko\v{s}ice, Ko\v{s}ice, Slovakia
\\
$^{118}$Technical University of Split FESB, Split, Croatia
\\
$^{119}$The Henryk Niewodniczanski Institute of Nuclear Physics, Polish Academy of Sciences, Cracow, Poland
\\
$^{120}$The University of Texas at Austin, Physics Department, Austin, Texas, United States
\\
$^{121}$Universidad Aut\'{o}noma de Sinaloa, Culiac\'{a}n, Mexico
\\
$^{122}$Universidade de S\~{a}o Paulo (USP), S\~{a}o Paulo, Brazil
\\
$^{123}$Universidade Estadual de Campinas (UNICAMP), Campinas, Brazil
\\
$^{124}$Universidade Federal do ABC, Santo Andre, Brazil
\\
$^{125}$University of Houston, Houston, Texas, United States
\\
$^{126}$University of Jyv\"{a}skyl\"{a}, Jyv\"{a}skyl\"{a}, Finland
\\
$^{127}$University of Liverpool, Liverpool, United Kingdom
\\
$^{128}$University of Tennessee, Knoxville, Tennessee, United States
\\
$^{129}$University of the Witwatersrand, Johannesburg, South Africa
\\
$^{130}$University of Tokyo, Tokyo, Japan
\\
$^{131}$University of Tsukuba, Tsukuba, Japan
\\
$^{132}$University of Zagreb, Zagreb, Croatia
\\
$^{133}$Universit\'{e} de Lyon, Universit\'{e} Lyon 1, CNRS/IN2P3, IPN-Lyon, Villeurbanne, Lyon, France
\\
$^{134}$Universit\`{a} di Brescia, Brescia, Italy
\\
$^{135}$V.~Fock Institute for Physics, St. Petersburg State University, St. Petersburg, Russia
\\
$^{136}$Variable Energy Cyclotron Centre, Kolkata, India
\\
$^{137}$Warsaw University of Technology, Warsaw, Poland
\\
$^{138}$Wayne State University, Detroit, Michigan, United States
\\
$^{139}$Wigner Research Centre for Physics, Hungarian Academy of Sciences, Budapest, Hungary
\\
$^{140}$Yale University, New Haven, Connecticut, United States
\\
$^{141}$Yonsei University, Seoul, South Korea
\\
$^{142}$Zentrum f\"{u}r Technologietransfer und Telekommunikation (ZTT), Fachhochschule Worms, Worms, Germany
\endgroup